\documentclass[11pt,a4paper, preprint,  superscriptaddress]{revtex4}
\usepackage[utf8]{inputenc}
\usepackage{hyperref}
\usepackage{pdflscape}
 \usepackage{amsmath}
\usepackage{amssymb}
 \usepackage{subfigure}
 \usepackage{epsfig}
\usepackage{amsmath,amsfonts}
\usepackage{graphicx}

\begin{document}
	\title{Thermodynamics of galaxy clusters in modified Newtonian potential}
	\author {Abdul W. Khanday}
	\email{abdulwakeelkhanday@gmail.com}
	\affiliation{{Department of Physics, National Institute of Technology  Srinagar, Jammu and Kashmir -190006, India.}}
\author {Sudhaker Upadhyay}
	\email{sudhakerupadhyay@gmail.com}
	
	\affiliation{Department of Physics, K. L. S. College, Nawada, Bihar 805110, India}
\affiliation{Department of Physics, Magadh University, Bodh Gaya,
 Bihar  824234, India}
	\affiliation{Inter-University Centre for Astronomy and Astrophysics (IUCAA) Pune, Maharashtra-411007 }
 \affiliation{School of Physics, Damghan University, Damghan, 3671641167, Iran}
	
		\author {Prince A. Ganai}
	\email{princeganai@nitsri.net}
	\affiliation{{Department of Physics, National Institute of Technology  Srinagar, Jammu and Kashmir-190006, India.}}

	\begin{abstract}
	We study the thermodynamics of galaxy clusters in a modified Newtonian potential motivated by a general solution to   Newton's ``sphere-point" equivalence theorem. We obtain the $N$ particle partition function by evaluating the configurational integral while accounting for the extended nature of galaxies (via the inclusion of the softening parameter $\epsilon$ into the potential energy function). This softening  parameter takes care of the galaxy-halos whose effect on  structuring the shape of
	the galactic disc has been found recently. The spatial distribution of the  particles (galaxies) is also studied in this framework. A comparison of the new clustering parameter $b_+$ to the original clustering parameters is presented in order to visualize the effect of the modified gravity. We also discuss the possibility of system symmetry breaking  via the behavior of the specific heat as a function of temperature. 
	\end{abstract}	
		\maketitle
\section{Introduction}
Like stars and galaxies, galaxy clusters are   one of  the fundamental building blocks of the universe. These systems form the topmost level of the hierarchical structure formation and possess the most extreme masses and dimensions of all  celestial objects. Due to the complex optical appearance, the systematic study of these systems hasonly recently begun.  Advanced observational techniques  have helped a lot in their proper understanding. Optical and X-ray observations have led to a deep understanding of the evolution and composition of these larger systems. The famous Sunyaev-Zeldovich (SZ) effect and the gravitational lensing effect have added important details to our knowledge on these systems. 
Observations and theoretical studies have led to the emergence of $\Lambda$ cold dark matter ($\Lambda$CDM) model as a standard cosmological model. This model is based on a spatially flat universe, a cosmological constant responsible for the accelerated expansion at late times and a structure seeded by quantum fluctuations at very early times during the period of inflation.

 ``Messier" was the first to list  103 nebulae in his catalog, 30 of which are now identified as galaxies.  Herschel was a pioneer in the study of large scale structure of universe.   Herschel classified around 2500 nebulae in his lifetime and recognized several groups of galaxies, such as Ursa Major, Leo, Hydra and NGC 4169 etc. In the mid 1920s  Hubble proved that the nebulae were extra-galactic in nature and bona fide galaxies like Milky way~\cite{hubble1925cepheids}. In 1927 Lundmark  started his investigation of the formation mechanism of galaxy clusters for the first time.   Zwicky~\cite{zwicky1933rotverschiebung}  pioneering work on estimating the mass of galaxy clusters and establishing the need for extra-matter, now known as dark matter (DM).  pioneered the work of the estimation of mass of galaxy clusters and established the need for extra-matter now known as dark matter (DM).

The theoretical models of cluster formation employ techniques focusing on any particular aspect of clustering that they aim to understand. The initial conditions, DM mass budget  and gravity determine many important bulk properties. The self-similar model by Kaiser is one such simple model \cite{kaiser1986evolution}. This model predicts cluster properties close to observational findings. The closed box nature of cluster potential provides ideal laboratory conditions to understand galaxy formation and    inside processes. 
 
The models that describe the modifications to the standard general relativity (GR) gravity have implications for cosmic evolution as well as that of density perturbations and, therefore, for cluster formation. For example, $f(R)$ models  which modify gravity by the inclusion of a general function $f(R)$ of the Ricci curvature scalar in the Einstein-Hilbert action enhance the gravitational forces on scales appropriate for structure formation~\cite{jk}.

The changes to the gravity law have prompted the study of the statistical and thermodynamic properties of galaxy clusters in light of the modified potentials \cite{5,6,7,8,9}. For instance, the effect of quantum correction on the thermodynamics of galaxy clusters were studied in ~\cite{upadhyay2017thermodynamics} and the effect of power-law modification to Newtonian potential on the spatial distribution function have been studied in Ref. \cite{aw}. Similarly, the  effect of the expansion of the universe on the clustering of galaxies has been discussed in Ref. \cite{hd} and the effect of dynamical dark energy on the galactic clustering has also been studied  \cite{de}. Also, the expansion effect of the universe on the galaxy clustering has  been investigated in Ref. ~\cite{eu}. Within these studies, it has been observed that the modified gravity law affects the thermodynamic and statistical properties of galaxy clusters considerably. 

In the roots of the Newton's gravity the theorem of ``sphere-point" equivalence lies. This principle helped Newton to relate the terrestrial gravity (apple falling towards earth) to the orbital motion of planets. We have realized Newtonian gravity as the weak-field limit of GR which enables communication between observations and prediction. It has been shown in Ref.~\cite{k2} that the force function satisfying Newton theorem is  given by
\begin{equation}
	F(r)= Ar^{-2}+ B r,\label{1}
\end{equation} 
where $A$ and $B$ are constants related to the gravitational constant $G$ and the cosmological constant $\Lambda$ 
 (within numerical factors), respectively. The force parameter $B$ intended to be   such that the additional force is attractive.  The weak-field GR thus ensures that any matter, seen or unseen [44], at large galactic scales   interacts by the law of Eq. (1).

  Any matter (seen or unseen) at large galactic scales interacts via the equation (\ref{1})~\cite{g2}. Equation (\ref{1}) defines a non-force-free field inside the spherical shell, thus contradicting the Newtonian concept of a force-free field inside. However, the non-force-free field inside the shell agrees well with the observation of the effect of the galactic halos on the structure of galactic disks~\cite{k}.
The second term in the equation (\ref{1}) corresponds to the cosmological constant term in the Newtonian formulation of the Friedmann cosmological equation~\cite{k2}.  Although there have been numerous studies on galaxy clustering under modified gravity, the most important approach of modified weak-field GR has been overlooked.  We try to bridge this gap. 

In this article, we   quantitatively study the effect of the extra term ($ r$ dependency) present in  Newton's law  on the statistical distribution of   galaxies. We also study the effects of this force form   on the thermodynamic equations of state, viz. free energy, entropy, and chemical potential, etc.   The effect of this correction on various thermodynamic quantities and the general distribution function is also   observed. We also study the behavior of specific heat as an indicator of phase transition.  The correction term  shows up in the form of a modified clustering parameter which estimates the extent of correlation among different system particles. 

\section{Partition function of the gravitating bodies under modified potential}
The force function given in  (\ref{1})  corresponds to a monogenic and conservative system and  related to the potential via the relation $F=-\frac{\partial \Phi}{\partial r}$. 
Utilizing the standard integral,  $\Phi =\int -F d R$, the potential energy function $\Phi$ takes the form ~\cite{ht}
\begin{equation}
\Phi(r) = -\frac{GM^2}{r}-Br^2,
\end{equation}
where $B=M\Lambda c^2/6$ (for numerical value see e.g. \cite{ht}).   
The partition function for  $N$-body system of point particles interacting gravitationally and each having mass $M$ and momenta $p_i$ is given by
~\cite{ahmad2002statistical}
\begin{equation}
Z_N\left(T,V\right)=\frac{1}{\lambda ^{3N} N!} \int d ^{3N}pd^{3N}r \exp\left(-\left[\sum_{i=1}^{N}\frac{p_i^2}{2M}+\Phi\left(r_1,r_2,...r_N\right)\right]T^{-1}\right),\label{3}
\end{equation}
where $\lambda$ is a normalization factor.

Performing the integration over  momentum space, we have
\begin{equation}
Z_N\left(T,V\right)=\frac{1}{N!}\left(\frac{2\pi MT}{\Lambda ^2}\right)I_N\left(T,V\right),\label{4}
\end{equation}
where  $I_N (T, V)$ is  configuration part of  integral (\ref{3}) and   given by
\begin{equation}
\begin{split}
I_N\left(T,V\right)&=\int.....\int\prod_{1\leq i <j\leq N}\exp(\left[-\Phi\left(r_i-r_j\right)T^{-1}\right])d^{3N}r,\\	
&=\int.....\int\prod_{1\leq i <j\leq N}\exp(\left[-\Phi\left(r_{ij}\right)T^{-1}\right])d^{3N}r.
\end{split}
\end{equation}
  The gravitational potential energy depends on the relative positions only, $ r_{ij}=r_i-r_j$, and, in general, estimated as the sum of potential energy of all pairs, i.e.
\begin{equation}
\Phi(r_1,r_2,...,r_N)=\sum_{1\leq i<j\leq N}\Phi(r_{ij}).
\end{equation}

The configuration integral can be solved by introducing a two-point function, $f_{ij}$~\cite{ahmad2002statistical}. Begin equation
	\begin{equation}
	f_{ij}=\exp\left(-\frac{\Phi_{ij}}{T}\right)-1,
	\end{equation} 
	to define the two-point function.
	The function  $f_{ij}$  vanishes   in the absence of interactions. The configuration integral takes the following form:
\begin{equation}
I_N\left(T,V\right)=\int....\int\left(1+f_{12}\right)\left(1+f_{13}\right)....\left(1+f_{N-1,N}\right)d^3r_1d^3r_2...d^3r_N.
\end{equation}
For point masses, the Hamiltonian function diverges at $r_{ij}=0$.   This discrepancy is eliminated by introducing a parameter $\epsilon$, which also accounts for the extended nature of galaxies, where  $0.01\leq\epsilon\leq0.05$ is measured in cell size units. \cite{ah}.  The interaction potential energy between galaxies takes on the form after the softening parameter $\epsilon$ is introduced   
\begin{equation}
\Phi =\left[\frac{-GM^2}{(r_{ij}^2+\epsilon^2)^{1/2}}-Br_{ij}^2\right].
\end{equation}
Because the second term does not diverge as $r_{ij}\rightarrow 0$, the softening parameter $\epsilon$ is not introduced.  The configuration integral for $N=1$, $I_1(T,V)$, takes the value 
\begin{equation}
I_1(T,V)=1.
\end{equation}
Begin equation 
\begin{equation}
I_2(T,V)=4\pi V\int_{0}^{R}\left[1+\frac{1}{(r^2 +\epsilon^2)^{(1/2)}T}+Br^2 \right]r^2d r,
\end{equation}
to calculate the configuration integral for $N = 2$.
Here we have used the fact that the effect of the long-range mean gravitational field is exactly balanced by the expansion of the universe~\cite{sh}. Upon solving, the above integral yields  
\begin{equation}
I_2(T,V)=V^2\Biggl[1+ \frac{3}{2} \frac{GM^2}{R T}\Biggl( \sqrt{1+\frac{\epsilon^2}{R ^2}}+\frac{\epsilon^2}{R ^2}\log\frac{\frac{\epsilon}{R }}{1+\sqrt{1+\frac{\epsilon^2}{R ^2}}} +\frac{2B}{5GM^2}R ^3\Biggr)\Biggr].  
\end{equation}
 In a  more compact form,  this equation can be written as
\begin{equation}
I_2(T,V)=V^2\left[1+\frac{3}{2}  \frac{GM^2}{R T} \left(C_1+C_2\right)\right],
\end{equation}
where 
\begin{align}
C_1&=\sqrt{1+\frac{\epsilon^2}{R ^2}}+\frac{\epsilon^2}{R ^2}\log\frac{\epsilon/R }{1+\sqrt{1+\frac{\epsilon^2}{R ^2}}},\notag\\
C_2&=\frac{2B R ^3}{5GM^2}.\notag
\end{align}
In general, the value of the configurational integral for higher values of $N$ is given by
\begin{equation}
I_N(T,V)=V^N\biggl[1+\left(C_1 +C_2 \right)\eta\biggr]^{N-1},\label{14}
\end{equation}
where 
\begin{equation}
\eta=\frac{3}{2}\frac{GM^2}{R T}.  
\end{equation}
Using equations (\ref{4}) and (\ref{14}), the grand partition function for the system of $N$ gravitationally interacting galaxies is given by
\begin{equation}
Z_N(T,V)=\frac{1}{N!}\left(\frac{2\pi MT}{\lambda^2}\right)^{3N/2}V^N\left[1+(C_1+C_2)\eta\right]^{N-1}.\label{15}
\end{equation}
This partition function includes the corrected parameter $C_2$ together with the uncorrected parameter  $C_1$.
\section{THERMODYNAMICS OF THE SYSTEM OF GALAXIES} 
Once the partition function is known, we can derive the thermodynamic equations of state for the system. 
The Helmholtz free energy carries all the useful information about the system that the partition function carries, and these are connected  by the relation $F(V,N,T)=-T\ln Z_N(T,V)$. For the system of galaxies, $F$ can be written as
\begin{equation}
F(V,N,T)=-T\ln\biggl[\frac{1}{N!}\left(\frac{2\pi MT}{\lambda^2}\right)^{3N/2}V^N\left[1+\left(C_1+C_2\right)\eta\right]^{N-1}\biggr].\label{16}
 \end{equation}
We approximate $N-1\approx N$ for the system of galaxies with a large $N$ value. With this approximation,  equation (\ref{16}) can be simplified to
\begin{eqnarray}
F=NT\left(\ln\frac{N}{V}T^{-3/2}\right)-NT-NT\ln\left[1+\left(C_1+C_2\right)\eta\right]-\frac{3}{2}NT\ln\biggl(\frac{2\pi M}{\lambda^2}\biggr).\label{17}
\end{eqnarray}
Here, we observe that    within the weak-field limit of GR
 the Helmholtz free energy of the system is modified by  term having parameter $C_2$ in the logarithmic term.

 The entropy of the system of galaxies is computed for  the above partition function  using the following relation:
\begin{equation}
S=T\biggl(\frac{\partial\log Z}{\partial T}\biggr)_{N,V}+\log Z.\label{18}
\end{equation}
Substituting the values from equation (\ref{15}) into equation (\ref{18}), the entropy of the system takes the following functional form:
\begin{eqnarray}
S&=&N\ln\left(\frac{V}{N}T^{3/2}\right)+N\ln[1+(C_1+C_2)\eta]-3N\frac{\left(C_1+C_2\right)\eta}{ 1+\left(C_1+C_2\right)\eta }\nonumber\\
&+&\frac{5}{2}N+\frac{3}{2}N\ln \frac{2\pi M}{\lambda^2}.\label{19}
\end{eqnarray}
This relationship  can be further simplified to
\begin{equation}
\frac{S}{N}=\ln\biggl(\frac{V}{N}T^{3/2}\biggl)-\ln[1-b_+]-3b_+
+\frac{5}{2}+\frac{3}{2}\ln \frac{2\pi M}{\lambda^2},
\end{equation}
where $b_+=\frac{\left(C_1+C_2\right)\eta}{ 1+\left(C_1+C_2\right)\eta }$ is a clustering parameter that takes values between $0$ and $1$ and estimates the strength of interaction. The Newtonian clustering parameter $b_epsilon$ defined in ~\cite{ahmad2002statistical} is related to the clustering parameter $b_+$ via the relationship
 \begin{equation}
	b_+= \frac{b_\epsilon (1-C_2\eta)+C_2\eta}{1+C_2\eta-b_\epsilon C_2\eta},
\end{equation} 
where $b_\epsilon=\frac{C_1\eta}{1+C_1\eta}$. In the above relation, when $C_2\rightarrow 0, b_+\rightarrow b_\epsilon$.
\par
\begin{figure}[h!]
	\centering
	\includegraphics[width=8 cm, height=6 cm]{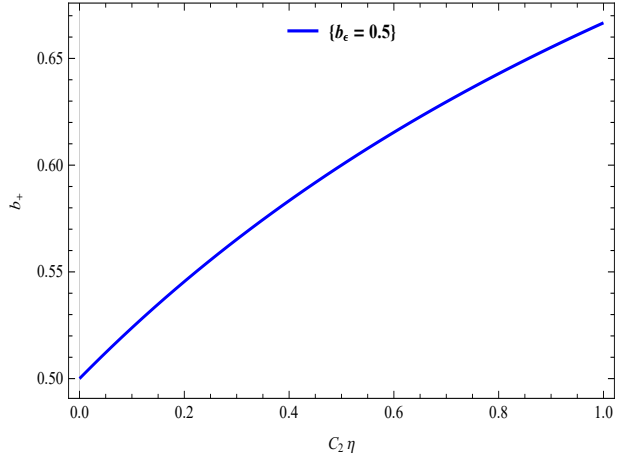}
	\caption{For a fixed value of the unmodified clustering parameter $b_\epsilon$, the figure depicts the variation of the modified clustering parameter $b_+$ with the strength of the correction term $C_2\eta$.}\label{fig1}
\end{figure}

 The variation of the clustering parameter $b_+$ with the strength of the correction term can be visualized from the  Fig. \ref{fig1}. The clustering becomes  stronger as the value of the correction term increases. 
\begin{figure}[h!]
	\centering
	\includegraphics[width=9 cm, height=6 cm]{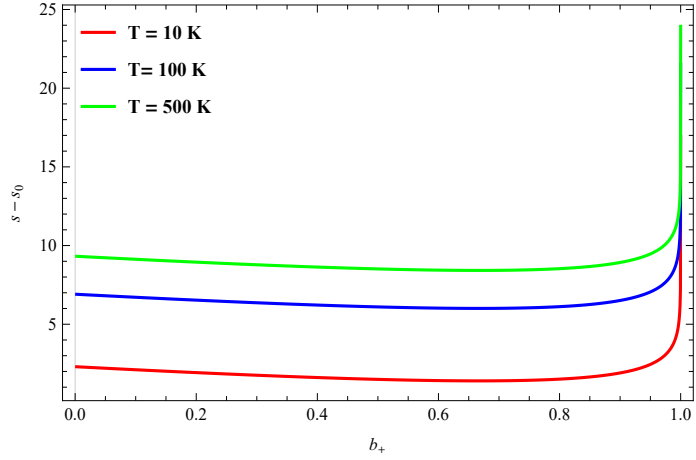}
	\caption{The graph depicts the isothermal variation of specific entropy $(S-S_0)$ as a function of clustering parameter $b_+$ for three different system temperatures and a fixed average density ($bar{n}$).} \label{fig2}
\end{figure}
\par
If   $\bar{n}T^{-3/2}$ is constant, the gravitational contribution to entropy $S-S_0=-\log[1-b_+]-3b_+$ reduces total entropy significantly for small $b_+$ but makes it infinite as $b \rightarrow 1$ due to a logarithmic divergence. This is  evidence of the undergoing phase-transition of the system as it becomes more and more clustered.  For $b=0$, the entropy  reduces to that of a non-interacting classical gas.
 \par 
Utilizing   free energy (\ref{17}) and entropy  (\ref{19}),  the internal energy for the system of galaxies  can be derived using the relation, $U=F+TS$, as
\begin{equation}
U=\frac{3}{2}NT\left[1-2b_+\right].\label{22}
\end{equation}

The pressure is related to Helmholtz free energy via the relation $P=-\left(\frac{\partial F}{\partial V}\right)_{N,T}$, which involves the first derivative of free energy w.r.t volume $V$, keeping particle number, $N$, and temperature, $T$, fixed. The pressure of the system of galaxies reads:  
\begin{equation}
P=\frac{NT}{V}\left[1-b_+\right]. 
\end{equation}

The chemical potential, $\mu$, measures the change in the internal energy of the system if a particle is added to the system while keeping volume and temperature constant, i.e.,  $\mu = \left(\frac{\partial F}{\partial N}\right)$. Using the relation for Helmholtz free energy,  the chemical potential takes the following functional form: 
\begin{equation}
\mu=T\left(\ln \frac{N}{V}T^{-3/2}\right)+T\ln \left[1-b_+\right]-Tb_+
-\frac{3}{2}T\ln(\frac{2\pi M}{\lambda^2}).\label{25}
\end{equation}
\par
\begin{figure}[ht]
	\centering
	\includegraphics[width=9 cm, height=6 cm]{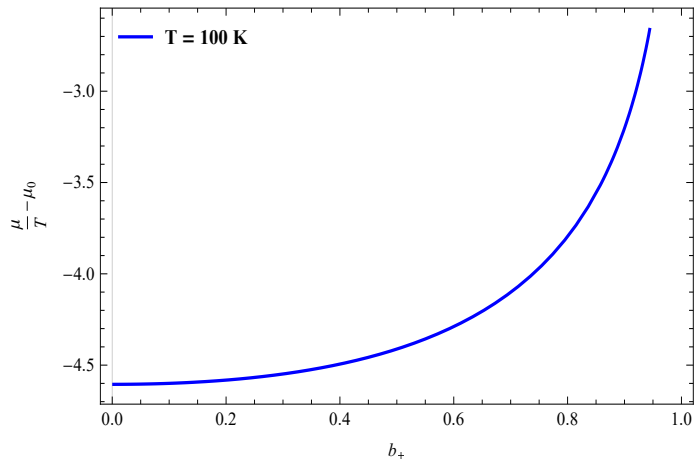}
	\caption{For a fixed value of $\bar{n}T^{3/2}$ , the figure depicts the variation of chemical potential  $\mu/T-\mu_0$  as a function of clustering parameter  $b_+$. The chemical potential is negative as the addition of a single particle decreases the total gravitational binding energy of the system.}\label{fig30}
\end{figure}

The above   chemical potential can further be expressed as
\begin{equation}
\frac{\mu}{T}-\mu_0=\left(\ln \frac{N}{V}T^{-3/2}\right)+\ln \left[1-b_+\right]-b_+.
\end{equation}
The behavior of the chemical potential   as a function of order parameter $b_+$ is shown in the following graph (Fig. \ref{fig30}). 

\section{General distribution function}
Now, we use the results derived above to obtain the functional form of the gravitational quasi-equilibrium distribution function (GQED), $F(N)$, for the cosmological many-body problem. We treat our system as an ensemble of sub-regions all of the same volume. We let the number of galaxies and their mutual gravitational energy fluctuate among the sub-systems. Thus, our system is a grand-canonical system characterized by a chemical potential $\mu$ and a temperature $T$.  The partition function (a weighted sum of all canonical partition functions) for the grand canonical ensemble is defined by 
\begin{equation}
Z_G(T,V,z)=\sum_{N=0}^{\infty} e^{\frac{N\mu}{T}}Z_N(T,V),\label{26}
\end{equation} 
 where   $z$  is the system activity given by    $z=\exp(\frac{\mu}{T})$.

The partition function for the system of galaxies interacting through the modified Newtonian potential is given by
\begin{equation}
\log Z_G=\frac{PV}{T}=N(1-b_+).
 \end{equation}

\begin{figure}[h!]
\begin{center}
 \includegraphics[width=.75\linewidth]{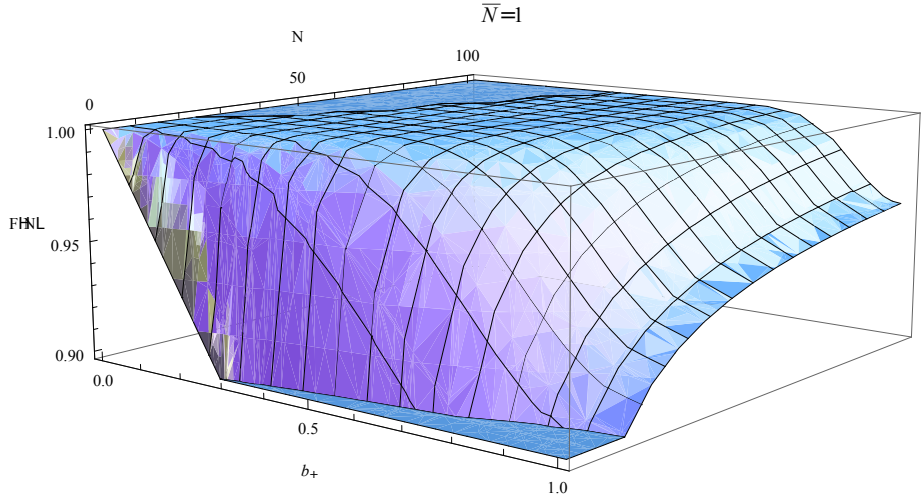}
 \caption{ For $\bar N=1$, the distribution function  vs. clustering parameter $b_+$.}
\label{fig01}
\end{center}
\end{figure}

\begin{figure}[h!]
\begin{center}
 {
\includegraphics[width=.75\linewidth]{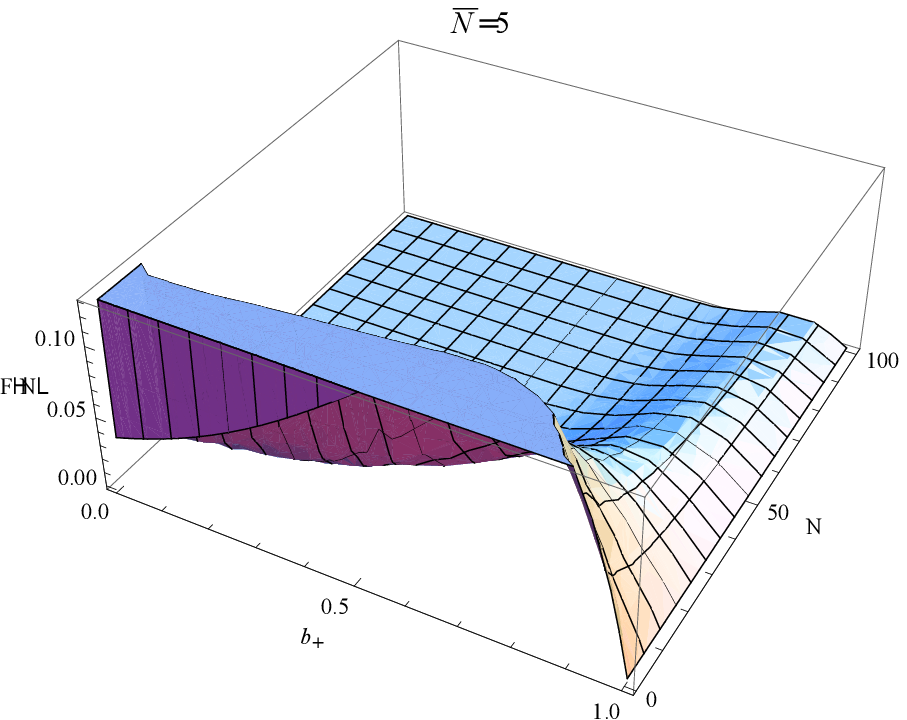}
\label{1b}}
 \caption{For $\bar N=5$, the distribution function  vs. clustering parameter $b_+$.}
\label{fig02}
\end{center}
\end{figure}

\begin{figure}[h!]
\begin{center}
 {\includegraphics[width=.75\linewidth]{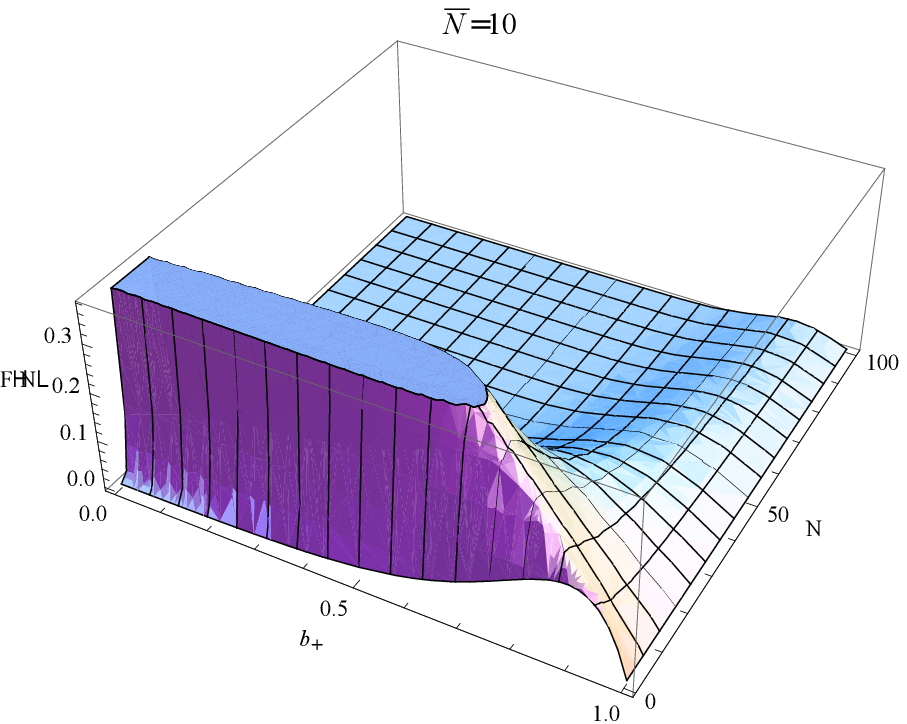}
\label{1c}}
 \caption{For $\bar N=10$, the distribution function  vs. clustering parameter $b_+$.}
\label{fig03}
\end{center}
\end{figure}

\begin{figure}[h!]
\begin{center}
 {
\includegraphics[width=.75\linewidth]{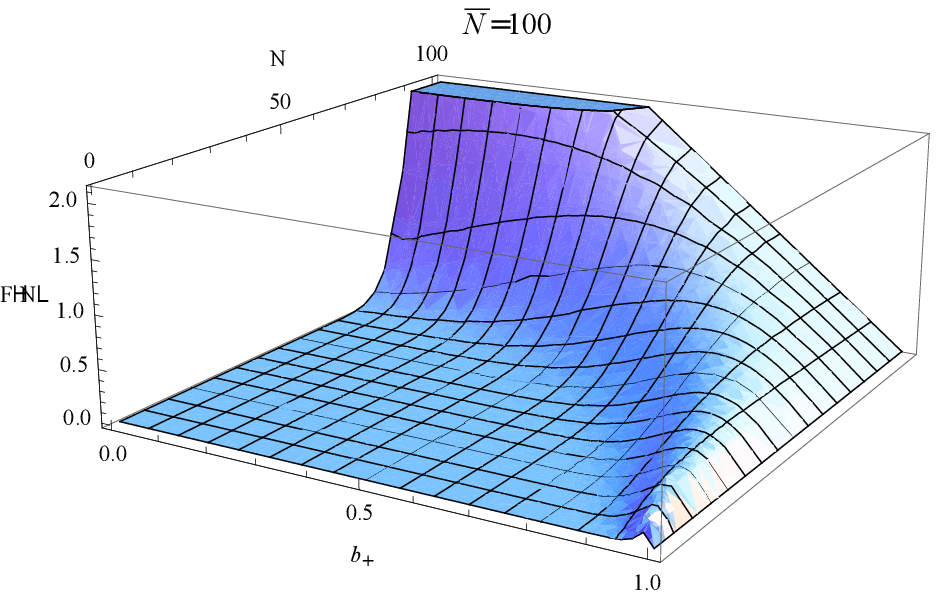}
\label{1d}}
\caption{For $\bar N=100$, the distribution function  vs. clustering parameter $b_+$.}
\label{fig4}
\end{center}
\end{figure}

The probability of finding $N$ particles in volume $V$ is calculated via the relation 
\begin{equation}
F(N)=\frac{\sum_{i}^{} e^{\frac{N\mu}{T}}e^{\frac{U_i}{T}}}{Z_G(T,V,z)}=\frac{e^{\frac{N\mu}{T}}Z_N(T,V)}{Z_G(T,V,z)}.\label{28}
\end{equation}
Using equations (\ref{15}), (\ref{26}), and (\ref{28}), the distribution function for the system of galaxies is given as
\begin{equation}
F(N)=\frac{\bar{N}}{N!}(1-b_+)[\bar{N}(1-b_+)+Nb_+]^{N-1}e^{-Nb_+-\bar{N}(1-b_+)}.\label{29}
\end{equation}
According to Ref. ~\cite{cs},  this thermodynamic distribution function agrees well with the Zwicky catalogue distribution of galaxies, and the clustering parameter is   $b_+\approx 0.75$. A more recent cluster catalogue described in   ~\cite{wh}  based on the Sloan Digital Sky Survey 3, and containing 132, 684 galaxy clusters in the $0.05\leq z\leq 0.8$ redshift range, also satisfies the distribution function described in equation (\ref{29}).    The distribution function $F(N)$ has a Poisson component modified by the corrected clustering parameter $b_+$. The pure Poisson distribution is restored in the limit  $b_+\rightarrow0$.
Remarkably, the basic structure of the distribution function is similar to that obtained by Saslaw and Hamilton ~\cite{saslaw1984thermodynamics}. The shapes of the distribution function as a function of particle number $N$ and clustering parameter $b_+$ are depicted in the figures
  \ref{fig01}, \ref{fig02}, \ref{fig03}, and \ref{fig4}.
  


\par
\begin{figure}[h!]
	\centering
	\includegraphics[width=8 cm, height=5 cm]{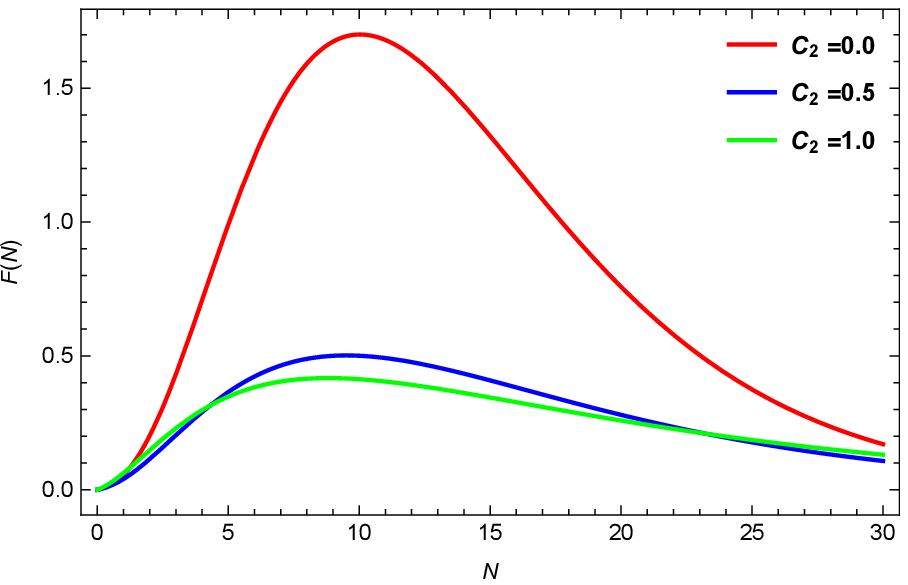}
	\caption{The distribution function $F(N)$'s behaviour as a function of particle number $N$. The red, blue, and green curves represent three different values for the correction parameter $C_2$. The red curve represents no correction, the blue curve represents $C_2=0.5$, and the green curve represents $C_2=1$. }
	\label{fig5}
\end{figure}

 {The behavior of the distribution function in two-dimensions  is also shown in   \ref{fig5}. From the graph, it can be seen that the   distribution function flattens as the value of the correction factor is increased.  From the plot, we also deduce that    the correction parameter shifts the peak downwards without changing the basic structure of the curve.}
\section{Power law for two-point correlation function}
 Large scale surveys have shown that galaxies are not uniformly distributed in the universe ~\cite{h}. Galaxies cluster on the scales of Mpc as well as in large-scale diverse structures on the scales of tens of  Mpc. To describe clustering at a large scale, two types of statistics are commonly used. One is the two-point correlation function $(\xi(r))$, and the other is the power spectrum of object distribution, which is described as the correlation function's Fourier transform.

Here, we analyze the behavior of the correlation function in the presence of a corrected Hamiltonian. The assumption made by  Peebles about the power-law form of the correlation function, which describes the probability of finding another object within a given radius  when the first object is identified  ~\cite{peebles1980large}, has been confirmed through $N$-body computer simulations ~\cite{itoh1993gravitational}.

 The effect of modified potential on the power-law form  of the correlation function can be visualized by writing the clustering parameter $b_+$ in the following form ~\cite{iqbal2006gravitational}:
\begin{equation}
b_+=\frac{\bar{n}}{6T}\int \left[\frac{GM^2}{(r^2+\epsilon^2)^{1/2}}+ Br^2\right]\xi_2(n,R,T) d{V}, 
\end{equation}
where $\bar{n}$ is the particle number density. 
\par
When the above equation is differentiated with respect to $V$, it yields 
\begin{eqnarray}
\frac{\partial b_+ }{\partial V}  &=&\frac{ \bar{n}}{6T}\frac{\partial}{\partial V} \int \left[\frac{GM^2}{(r^2+\epsilon^2)^{1/2}}+Br^2\right]\xi_2(n,R,T) d{V}\nonumber\\
&+&\frac{1}{6T}\left(-\frac{\partial \bar{\rho}}{\partial V} \right)\int \left[\frac{1}{(r^2+\epsilon^2)^{1/2}}+ Br^2\right]\xi_2(n,R,T) d{V},
\end{eqnarray}
where we have used $\frac{\partial V}{\partial \bar{n}} =-\frac{\bar{n}}{V}$. 

Using the relation $\frac{\partial b_+ } {\partial \bar{n}}=\frac{b_+(1-b_+)}{\bar{n}}$ in the  preceding equation, we obtain the following relation for the power-law (for extended galaxies):
\begin{equation}
\xi_2(r)=\frac{9Tb_+^2}{2\pi  \bar{n} }\left[\frac{GM^2}{(r^2+\epsilon^2)^{1/2}}+Br^2\right].
\end{equation}
If we neglect the extended nature of galaxies  and  treat  them as point particles, the above  equation  takes the form
\begin{equation}
\xi_2(r)=\frac{9Tb_+^2}{2\pi  \bar{n} }\left[\frac{1}{r}+Br^2\right].
\end{equation}
\section{Phase Transition and the behavior of specific heat}
The initial Poisson distribution (zero correlation) of a gravitationally driven many-body system evolves through different stages. The evolution can be described  as the phase transition from uncorrelated phase to some positive value of correlation characterized by an order parameter $b_+$, i.e., $b_+=0$ to $b_+>0 $. As clustering enhances via gravitational interaction, the homogeneity of the system is lost and lumps are created. The behavior of specific heat is an important indicator of a phase transition. We analyze the variation of the specific heat with the growth in the system temperature.  

 The specific heat at constant volume is defined as
\begin{equation}
	C_V=\frac{1}{N}\left(\frac{\partial U}{\partial T}\right)_{V,N}. \label{34}
\end{equation}
Using equations (\ref{34}) and (\ref{22}), the expression for specific heat is
\begin{equation}
	C_V=\frac{3}{2}\left[\frac{1+6C\eta -4C^2\eta^2}{\left(1+C\eta\right)^2}\right],\label{35}
\end{equation} 
where $C=C_1+C_2$ has been substituted.

Initially, when there is no clustering i.e. $b_+=0$, the value of specific heat $C_V=3/2$. When the system becomes fully virialized i.e $b_+=1$, $C_V=-3/2$. In between these two extreme values there lies a critical value of temperature, $T=T_C$ at which $C_V$ takes the extremum (maximum) value 
\begin{equation*}
	\frac{\partial C_V}{\partial T}=0.
\end{equation*} 
The value of the critical temperature for the system of galaxies interacting via the corrected Newtonian potential is given by
\begin{equation}
	T_C=\left[3\frac{\bar{N}}{V}\left(GM^2\right)^3\left(C_1+C_2\right)\right]^{1/3}.
\end{equation}

In terms of critical temperature, the specific heat $C_V$  (\ref{35})  can be expressed as follows: 
\begin{equation}
 C_V=\frac{3}{2}\left[1-2\frac{1-4\left(T/T_C\right)^3}{\left\{1+2\left(T/T_C\right)^3\right\}^2}\right].
\end{equation}
\begin{figure}[ht]
	\centering
	\includegraphics[width=9 cm, height=6 cm]{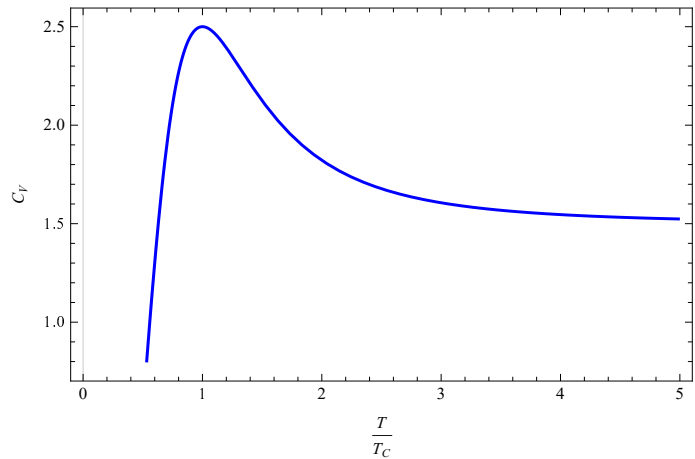}
	\caption{The graph depicts the change in specific heat $C_V$ as a function of temperature $T/T_C$. It can be  seen that the system symmetry breaks at $T=T_C$. }
	\label{fig6}
\end{figure}

 {Figure \ref{fig6} shows that at $T=T_C$, $C_V=5/2$, which is a typical diatomic gas property. At this point, the system homogeneity breaks at the average inter-particle separation by the formation of bound binary systems.  This marks the initiation of hierarchical phase transition at the lowest possible scale. The system symmetry breaking occurs on increasing scales, which is in principle different from the spontaneous phase transition commonly known.}

\section{Conclusion}
A sphere of mass   $M$  gravitates as a point particle of mass  $M$ located at its center, according to Newton's ``sphere-point" equivalence theorem, which connects terrestrial gravity to the orbital motion of planets in the solar system. The general solution that satisfies this theorem, $f(r)= Ar^{-2}+Br$, is the low energy limit of GR with a positive value of the cosmological constant. This connection is evident through the correspondence of the current observed value of the cosmological constant with the  value of the cosmological term in the modified Newtonian gravity law.   
 In the present paper, we have presented a quantitative  study of the clustering of galaxies under this modified Newtonian gravity.
  It was observed that  the corrected Newtonian potential modifies the clustering parameter, which reflects the effect of the correction on the clustering.
   We have calculated the  partition function for the  gravitationally interacting system of  galaxies. We have used the grand partition function to calculate the various thermodynamic equations of state like Helmholtz free energy, entropy, pressure,  and chemical potential. We also studied the effect of the modified potential on the spatial distribution of the galaxies. The  correlation of the system particles is also studied under a modified Newtonian potential.  
  The evolution of the system as characterised by the specific heat behaviour indicates the possibility of a phase transition around the critical temperature $T_C$.

  \end{document}